\documentclass{article}
\usepackage{icannga05}
\usepackage{times}
\usepackage[english]{babel}

\usepackage{amsmath}
\usepackage{amssymb}
\usepackage[ansinew]{inputenc}
\usepackage{tabularx}
\usepackage{enumerate}
\usepackage{url}
\usepackage{subfigure}
\usepackage{pstricks}
\usepackage{epsfig}

\newcommand{\mb}[1]{\mathbb{#1}}

\newcommand{\mf}[1]{\mathfrak{#1}}
\newcommand{\mr}[1]{\mathrm{#1}}

\newcommand{\braket}[2]{\left \langle #1 \, \left | \right . \, #2 \right \rangle }

\newcommand{\ket}[1]{\left | \, #1 \right \rangle}

\newcommand{\oneI}{1\hspace{-0.1cm}{\rm I}}

\newcommand{\modulo}[1]{\left | #1 \right |}

\newcommand{\fracd}[2]{{\displaystyle \frac{#1}{#2}}}

\newcommand{\urlbib}[1]{{\footnotesize{\url{#1}}}}

\begin{document}

\title{Numerical Simulations of a Possible Hypercomputational Quantum Algorithm}


\author{Andrés Sicard, Juan Ospina, Mario Vélez
\\ 
Logic and Computation Group, EAFIT University, A.A. 3300, Medellín, Colombia
\\
E-mail: \{asicard, jospina, mvelez\}@eafit.edu.co
}  

\maketitle
\begin{abstract}
The hypercomputers compute functions or numbers, or more generally solve problems or carry out tasks, that cannot be computed or solved by a Turing machine. Several numerical simulations of a possible hypercomputational algorithm based on quantum computations previously constructed by the authors are presented. The hypercomputability of our algorithm is based on the fact that this algorithm could solve a classically non-computable decision problem, the Hilbert's tenth problem. The numerical simulations were realized for three types of Diophantine equations: with and without solutions in non-negative integers, and without solutions by way of various traditional mathematical packages. 
\end{abstract}

\section{Introduction}
Although there have been several conferences
and various specialized publications
dedicated to the topic of hypercomputation, this is still an area in its gestation and development process. The term `hypercomputer' denotes any data processing device (theoretical, potencially realizable or that can be implemented) capable of carrying out tasks that cannot be performed by a Turing machine (TM) \cite{Copeland-2002,Burgin-Klinger-2004}. Proposed models of hypercomputation surge from diverse disciplines, such as non-classical logics, computability, neural networks, or relativistic physics \cite{Copeland-2002,Stannett-2003a,Sicard-Velez-2001a-english}. Notwithstanding the proliferation of theoretical hypercomputation models, the possibility of a \emph{conceivable} construction of a hypermachine, continues to be controversial and held under study and analysis. 

This work presents several numerical simulations of a possible hypercomputational algorithm based on quantum computations previously constructed by the authors \cite{Sicard-Velez-Ospina-2004}. The hypercomputability of our algorithm is based on the fact that this algorithm could solve a TM non-computable decision problem, Hilbert's tenth problem \cite{Matiyasevich-1993}. Our algorithm is based on the algorithm proposed by Tien D. Kieu 
\cite{Kieu-2003c,Kieu-2003a,Kieu-2003b}, with the difference that
we have selected the Infinite Square Well (ISW) instead of the
(one-dimensional) Simple Harmonic Oscillator (SHO) as the
underlying physical system. Since our model exploits the quantum adiabatic process for unbounded operators \cite{Avron-Elgart-1999}, and the characteristics of the infinite-dimensional irreducible representation of the dynamical algebra $\mf{su}(1,1)$ associated to the ISW \cite{Antoine-Gazeaub-Monceauc-Klauder-Penson-2001}, the simulation is only able to partially illustrate our algorithm. In other words, this work does not pretend to affirm that the numerical simulations constitute our hypercomputational quantum algorithm. 

\section{The Hypercomputational Quantum Algorithm}
The Hamiltonian operator $H$  and the energy levels  $E_{n}$, for a particle with mass $m$ in an ISW with length $\pi a$ are \cite{Antoine-Gazeaub-Monceauc-Klauder-Penson-2001}
\begin{equation}\label{eq-10}
  \begin{split}
   H &= i^2 \fracd{\hbar^2}{2m}\fracd{d^2}{dx^2} - \fracd{\hbar^2}{2ma^2} ,
\\
E_n &= (\hbar^2/2ma^{2})n(n+2)  , 
\end{split}
\end{equation}
where the action of $H$ on its eigenvectors
\begin{equation}\label{eq-20}
\left \{ \ket{n}  \mid n \in \mb{N} \right \}, \quad \mb{N}=\{0, 1, \dots \} , 
\end{equation}
is given by $H \ket{n} = E_n \ket{n}$. Due to the spectral structure of the ISW, the dynamical algebra associated with it, is the Lie algebra $\mf{su}(1,1)$ \cite{Antoine-Gazeaub-Monceauc-Klauder-Penson-2001}. This is a tridimensional algebra that satisfies the commutation relations $[K_-,K_+] = K_3$ and $[K_{\pm},K_3] = \mp2K_{\pm}$, where operators $K_+, K_-$ and $K_3$ are called creation, annihilation and Cartan operators, respectively. The algebra $\mf{su}(1,1)$ admits an infinite-dimensional irreducible representation where actions of $K_+, K_-$ and $K_3$ on basis \eqref{eq-20} are
\begin{align*}
    K_+\ket{n} &= \sqrt{(n+1)(n+3)}\ket{n+1} , 
    \\
    K_-\ket{n} &= \sqrt{n(n+2)}\ket{n-1} , 
    \\
    K_3\ket{n} &= (2n+3)\ket{n} . 
\end{align*}

With basis on the algebra $\mf{su}(1,1)$, the Hamiltonian \eqref{eq-10} is rewritten as $H = (\hbar^2/2ma^{2})K_+K_-$, and a number operator $N$ is given by
\begin{equation}\label{eq-70} 
N = (1/2)(K_3 - 3) .  
\end{equation}

Due to the dynamical algebra associated, the Barut-Girardello coherent states $\ket{z}, z \in \mb{C}$, for the ISW are eigenstates of annihilation operator $K_-$ \cite{Wang-Sanders-Pan-2000}
\begin{equation*}
\ket{z} = \frac{\modulo{z}}{\sqrt{I_2(2\modulo{z})}}\sum_{n=0}^{\infty}\frac{z^n}{\sqrt{n!(n+2)!}}\ket{n}  .
\end{equation*}

On the other hand, a Diophantine equation is of the following form
\begin{equation}\label{eq-90}
D(x_1, \dots, x_k) = 0  ,   
\end{equation}
where $D$ is a polynomial with integer coefficients. In present terminology, Hilbert's tenth problem may be paraphrased as: Given a Diophantine equation of type \eqref{eq-90}, we should build a procedure to determine whether or not this equation has a solution in non-negative integers. From the concluding results obtained by Matiyasevich, Davis, Robinson, and Putnam we know that, in the general case, this problem is algorithmically insolvable or more precisely, it is TM incomputable \cite{Matiyasevich-1993}.

With the basis for the ISW, our hypercomputational quantum algorithm for the Hilbert's tenth problem \cite{Sicard-Velez-Ospina-2004} is shown in Table \eqref{tabla-10}. Our algorithm is probabilistic such are quantum algorithms in general. Unlike these however, our algorithm is of adiabatic quantum computation over infinite-dimensional spaces, given that parting from initial ground state \eqref{eq-100} of \eqref{eq-120}, the final ground state \eqref{eq-145} of \eqref{eq-130}, is obtained via the adiabatic theorem for unbounded operators  \cite{Avron-Elgart-1999}. As we have stated elsewhere \cite{Sicard-Velez-Ospina-2004}, the construction of our algorithm, breaking from Kieu's algorithm, opens the possibility to construct similar algorithms based on physical references whose associated dynamic algebra possesses an infinite-dimensional irreducible representation.
\begin{table}
\caption{Hypercomputational quantum algorithm.}
\small
\begin{tabularx}{\linewidth}{|X|}
\hline
Given a Diophantine equation with $k$ unknowns of type \eqref{eq-90}, we provide the following quantum algorithm to decide whether this equation has any non-negative integer solution or not:
\begin{enumerate}[1.]
\item Construct a physical process in which a system initially
starts with a direct product of $k$ coherent states $\ket{\psi(0)}$, and in which the system is subject to a time-dependent Hamiltonian $H_{\mr{A}}(t)$ over the
time interval $[0,T]$, for some time $T$, with the initial Hamiltonian $H_{\mr{I}}$ and the final Hamiltonian $H_{\mr{D}}$, given by
{\begin{align}
\ket{\psi(0)} &= \bigotimes_{i=1}^k \ket{z_i}  , \label{eq-100}
\\
H_{\mr{A}}(t) &= (1 - t/T)H_{\mr{I}} +  ( t/T) H_{\mr{D}} , \nonumber 
\\
H_{\mr{I}} &= \sum_{i=1}^k \left(K_{+_i} - z^*_i\right)
\left(K_{-i} - z_i\right) , \label{eq-120}
\\
H_{\mr{D}} &= \left(D(N_1, \ldots, N_k)\right)^2 . \label{eq-130}
\end{align}}
\item Measure through the time-dependent Schr\"odinger equation with the 
{\begin{equation}\label{eq-135}
  i\partial_t \ket{\psi(t)} = H_{\mr{A}}(t) \ket{\psi(t)}, \mbox{ for $t \in [0,T]$}
\end{equation}}
the maximum probability to find the system in a
particular number state at the chosen time $T$,
{\begin{align}
    \begin{split}\label{eq-140}
P_{\mr{max}}(T) &= \max_{(n_1,\dots,n_k) \in \mb{N}^k} \left | \braket{\psi(T)}{n_1,\dots,n_k}  \right |^2 ,
\\
&=\left | \braket{\psi(T)}{n_1,\dots,n_k}_0 \right | ^2 ,
\end{split}
\end{align}}
where
{\begin{equation}\label{eq-145}
  \ket{\{n\}}_0 \equiv \ket{n_1,\dots,n_k}_0  .
\end{equation}}

\item If $P_{\mr{max}}(T) \le 1/2$, increase $T$ and repeat all the steps above.
\item If 
  \begin{equation}\label{eq-150}
    P_{\mr{max}}(T)>1/2     
  \end{equation}
then $\ket{\{n\}}_0$ is the ground state of $H_{\mr{D}}$ 
(assuming
no degeneracy) and we can terminate the
algorithm and deduce a conclusion from the fact that
\begin{equation}\label{eq-160}
H_{\mr{D}}\ket{\{n\}}_0 = 0 \mbox{ iff  \eqref{eq-90} has a 
solution in $\mb{N}$}  .  
\end{equation}
\end{enumerate}
\\
\hline
\end{tabularx}
\label{tabla-10}
\end{table}

\section{Simulation Procedure}
Once constructed  \eqref{eq-130} with \eqref{eq-100} and \eqref{eq-120} selected for a Diophantine equation of type \eqref{eq-90}, our problem revolves around solving  \eqref{eq-135} in search of a value of $T$ which satisfies condition \eqref{eq-150}, whereby we can establish criteria \eqref{eq-160}. Given that  \eqref{eq-135} is formulated inside an infinite dimensional Fock space, the simulation consists in being solved numerically in a Fock space truncated from dimension $m+1$, selected such that  \cite{Kieu-2003} 
\begin{equation*}
\ket{z}^{m+1} =
\frac{\modulo{z}}{\sqrt{I_2(2\modulo{z})}}\sum_{n=0}^{m}\frac{{z}^n}{\sqrt{n!(n+2)!}}\ket{n}^m 
\end{equation*}
has a norm less than one by some chosen $\epsilon$
\begin{equation}\label{eq-180}
  \left | 1- \sqrt{^{m+1}\braket{z}{z}^{m+1}} \right | \le\epsilon ,
\end{equation}
where $\ket{z}^{m+1}$ represents the infinite-dimensional state  $\ket{z}$ truncate to the  $m+1$ dimension. 
 
The numerical solution involves a discretization procedure
that consists in replacing \eqref{eq-135}
by a discrete equation, whose solver preserves the norm of the state
vector in the evolution of time. This solver is unitary and it is
given by the Cayley transform of $H_A^{m+1}$ \cite{Kieu-2003}
\begin{equation}\label{eq-190}
\ket{\psi(t+1)}^{m+1} =
\frac{1-\frac{i}{2}H_A^{m+1}(t)} {1 +
\frac{i}{2}H_A^{m+1}(t)} \ket{\psi(t)}^{m+1}  .
\end{equation}

\section{Simulation parameters and results}
For the numerical simulations we have selected three Diophantine equations. Since we counted on all the information available for the simulation, the maximum probabilities  \eqref{eq-140} can be calculated directly.  Of course, this situation would be different if we were measuring upon the physical reference, in which case these probabilities should be obtained parting from the expected values of the operator numbers \eqref{eq-70}. 

\subsection{Equation with solution}
We now consider a simple example which nevertheless has all the
interesting ingredients typical for a general simulation of our
hypercomputing  quantum algorithm. Let $D(x)$ be the Diophantine equation
\begin{equation}\label{eq-200}
 D(x) \equiv x-6 = 0  .
\end{equation}

In relation to the values of $z$, these should be greater than $1.6$ to guarantee that the maximum peak of density of probability associated with the initial ground state  \eqref{eq-100} is less than $1/2$ \cite{Sicard-Velez-Ospina-2004}. We choose $z = 4$ and $m=9$ which satisfy  \eqref{eq-180} for $\epsilon = 10^{-4}$. The truncated Fock space has only ten dimensions, and is
generated by $\{\ket{0},\dots,\ket{9}\}$. For \eqref{eq-200}, according to the algorithm pointed out in Table  \eqref{tabla-10}, we have that
\begin{align*}
& \ket{\psi(0)}^{10} \approx 
\\
& [0.16 \; 0.36 \; 0.51 \; 0.53 \; 0.43 \; 0.29  \; 0.17 \; 0.08 \; 0.04 \; 0.02 ]^T ,
\\
&  H_{\mr{D}}^{10 \times 10} =  \left( N^{10 \times 10} - 6\oneI^{10 \times 10} \right)^2  ,
\\
& H_{\mr{I}}^{10 \times 10} = \left(K_+^{10 \times 10} - 4\oneI^{10 \times 10}\right) \left(K_-^{10 \times 10} - 4\oneI^{10 \times 10}\right ).  
\end{align*}

\begin{figure}[ht]
     \epsfig{figure=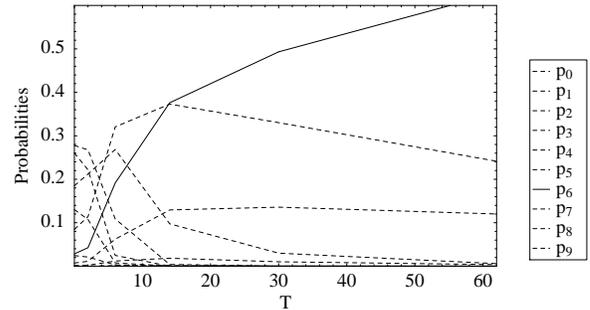,scale=.8}
     \caption{Simulation results for \eqref{eq-200} with $(z,m)=(4,9)$.}
     \label{fig-10}
 \end{figure}

Applying \eqref{eq-190} and incrementing $T$, figure \eqref{fig-10} indicates that the maximum probability reaches the value of $1/2$ for $T \approx 35$ and then
the Fock state $\ket{6}^{10} = \begin{bmatrix} 0 & \dots & 1 & 0 & 0 & 0 \end{bmatrix}^T$,  
is identified as the final ground state \eqref{eq-145} according to  \eqref{eq-150}. Now, given that  $H_{\mr{D}}^{10 \times 10} \ket{6}^{10} = 0$, we can conclude that \eqref{eq-200} is solved in the non-negative integers in agreement with  \eqref{eq-160}.

\subsection{Equation without solution}
\begin{figure}[ht]
   \begin{center}
     \epsfig{figure=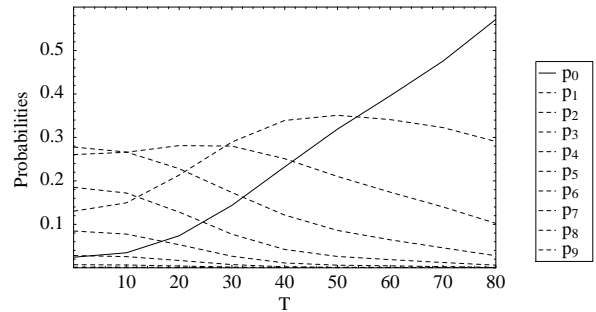,scale=0.8}
     \caption{Simulation results for \eqref{eq-210} with $(z,m)=(4,9)$.}
     \label{fig-20}
   \end{center}
 \end{figure}

Let $D(x)$ be the Diophantine Equation
 \begin{equation}\label{eq-210}
  D(x) \equiv x+6 = 0 , 
 \end{equation}
and let $z=4$ and $m=9$. The figure \eqref{fig-20} shows the probabilities of states as a function of
$T$. Below $T \approx 70$ none of the states have probabilities greater than $1/2$, and in fact the excited states, clearly
dominate in this regime. Eventually, we enter the quantum adiabatic
regime upon which the dominant state raises its probability over the
$1/2$ value; indeed it corresponds to the Fock state $\ket{0}$. Parting from \eqref{eq-160}, it is concluded then that \eqref{eq-210} cannot be solved in the non-negative integers.

Unlike the previous example, this simulation does not give a conclusive result. In fact, it would be more precise to affirm that according to figure \eqref{eq-20}, equation \eqref{eq-210} does not have a solution in  $\{0, \dots, m=9 \}$. Therefore, it would be necessary to increase the value of $m$ and execute the algorithm once again. Unfortunately, different to what would occur if we could count on the physical reference, in the case of the simulation, a standing criteria to determine the maximum value of m from which we can obtain a conclusive result, does not exist.

\subsection{Catalan  Equation} 
\begin{figure}[ht]
   \begin{center}
     \epsfig{figure=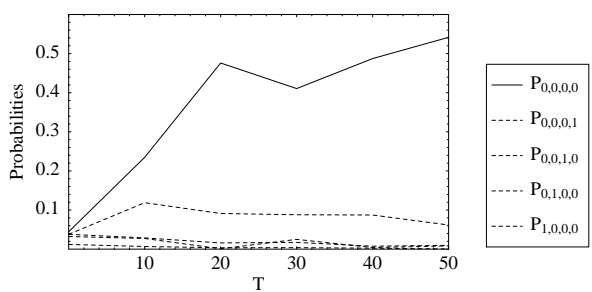,scale=0.8}
     \caption{Simulation results for \eqref{eq-220} with $(z,m)=(1.6,3)$.}
     \label{fig-30}
   \end{center}
 \end{figure}

Given that every exponential Diophantine equation can be reduced to a polynomial Diophantine equation  \cite{Matiyasevich-1993}, Hilbert's tenth problem also applies to these types of equations. It was recently proven that the Catalan Equation $x^p - y^q = 1$, with $x,y,p,q \geq 2$,
has one and only one
solution $(x=q=3,y=p=2)$ \cite{Waldschmidt-2004}. Although this solution could be found by simple inspection, different mathematical packages such as \emph{Mathematica$^{\small TM}$} or  \emph{Maple$^{\small TM}$}, cannot find it. To put our algorithm to work in order to determine if the Catalan Equation can be solved in non-negative integers, we rewrite it as 
\begin{equation}\label{eq-220}
  (a+3)^{(b+2)} - (c+2)^{(d+3)} = 1 ,
\end{equation}
whose only solution is $(a=b=c=d=0)$. The simulation results obtained are shown in figure \eqref{fig-30}. Upon verification of  \eqref{eq-160}, we can observe that our algorithm determines the existence of a solution.

\section{Conclusions}
Although the numerical simulations cannot be equivalent to our possible hypercomputational quantum algorithm, due to the impossibility to simulate an infinite
number of dimensions, their results give an idea of the behaviour that could be observed if we would have a real implementation and execution of our algorithm, supported in a determined physical process. 

The nonequivalence is observed through an asymmetry between solvable and non-solvable Diophantine Equations in the non-negative integers, with respect to the conclusiveness of the numerical simulations. This asymmetry is usually a common characteristic of the Turing machine incomputable problems, for which, if the instance of the problem has a solution, a classic algorithm would find it; while if it doesn't have a solution, such algorithm will not halt.

\section*{Acknowledgments} 
We thank Tien D. Kieu for helpful discussions and feedback. This work was supported by COLCIENCIAS (grant \#RC-284-2003) and by EAFIT University (grant \#1216-05-13576).


\end{document}